\documentclass[life,article,submit,moreauthors,pdftex,10pt,a4paper]{Definitions/mdpi} 

\firstpage{1} 
\pubvolume{xx}
\issuenum{1}
\articlenumber{5}
\pubyear{2018}
\copyrightyear{2018}
\history{Received: date; Accepted: date; Published: date}

\usepackage{amssymb}
\usepackage[version=4]{mhchem}
\usepackage[flushleft]{threeparttable}

\definecolor{DarkGreen}{RGB}{17,69,17}

\newcommand{\sed}[1]{#1}

\newcommand{\rrsed}[1]{\textbf{\textcolor{DarkGreen}{#1}}}

\Title{Origin of life's building blocks in Carbon and Nitrogen rich surface hydrothermal vents}

\Author{Paul B. Rimmer $^{1,2,3,*}$\orcidA{} and Oliver Shorttle $^{1,4,}$\orcidB{}}

\AuthorNames{Paul B. Rimmer and Oliver Shorttle}

\address{%
$^{1}$ \quad Department of Earth Science, University of Cambridge, Downing Street, Cambridge, CB2 3EQ, United Kingdom\\
$^{2}$ \quad Cavendish Astrophysics, University of Cambridge, JJ Thomson Ave, Cambridge, CB3 0HE, United Kingdom\\
$^{3}$ \quad MRC Laboratory of Molecular Biology, Francis Crick Ave, Cambridge, CB2 0QH, United Kingdom\\
$^{4}$ \quad Institute of Astronomy, Madingly Road, Cambridge, CB3 0HA, United Kingdom}

\corres{Correspondence: pbr27@cam.ac.uk; Tel.: +44 (0) 1223 333 462}

\abstract{There are two dominant and contrasting classes of origin of life scenarios: those predicting that life emerged in submarine hydrothermal systems, where chemical disequilibrium can provide an energy source for nascent life; and those predicting that life emerged within subaerial environments, where UV catalysis of reactions may occur to form the building blocks of life.  Here, we describe a prebiotically plausible environment that draws on the strengths of both scenarios: surface hydrothermal vents. We show how key feedstock molecules for prebiotic chemistry can be produced in abundance in shallow and surficial hydrothermal systems. We calculate the chemistry of volcanic gases feeding these vents over a range of pressures and basalt C/N/O contents. If ultra-reducing carbon-rich nitrogen-rich gases interact with subsurface water at a volcanic vent they result in $\text{10}^\text{-3}$ -- $\text{1}\,\text{M}$ concentrations of diacetylene (\ce{C_4H_2}), acetylene (\ce{C_2H_2}), cyanoacetylene (\ce{HC_3N}), hydrogen cyanide (\ce{HCN}), bisulfite (likely in the form of salts containing \ce{HSO_3^-}), hydrogen sulfide (\ce{HS^-}) and soluble iron in vent water. One key feedstock molecule, cyanamide (\ce{CH_2N_2}), is not formed in significant quantities within this scenario, suggesting that it may need to be delivered exogenously, or formed from hydrogen cyanide either via organometallic compounds, or by some as yet-unknown chemical synthesis. Given the likely ubiquity of surface hydrothermal vents on young, hot, terrestrial planets, these results identify a  prebiotically plausible local geochemical environment, which is also amenable to future lab-based simulation.}


\begin{document}


\section{Introduction}
\label{sec:introduction}

In the past twenty years, significant progress has been made in determining how the building blocks of life can be produced.  The most successful model thus far considers the role of photochemistry, building larger molecules from an initial set including hydrogen cyanide and cyanoacetylene \cite{Powner2009,Ritson2012,Patel2015,Xu2018}.  This scenario centers around the photodetachment of an electron from an anion, which then reduces the hydrogen cyanide, leading to simple sugars which, in the presence of cyanamide, cyanoacetylene and phosphates, produce pyrimidine nucleotides: the building blocks of RNA. Reactions with ammonia or hydrolysis at key stages diverges the chemistry, resulting in amino acids (the building blocks of proteins) and lipids (the building blocks of cell membranes). The best anion thus far identified for this chemistry is bisulfite \cite{Xu2018}. The environment in which these prebiotic reactions can take place requires high concentrations of reactants, wet-dry cycles for phosphorylation, and exposure to ultraviolet light between 200 and 280 nm. These conditions strongly favor a surface environment. 

In contrast, compelling evidence that early life resided in hydrothermal vents has lead to speculation that life also originated within these deep sea environments. Phylogenetic studies point towards life's last universal common ancestor occupying a vent habitat \cite{Weiss2016}; vents possess natural pH, thermal, and chemical potential gradients that life can exploit for energy \cite{Russell2004,Russell2010};  and there are chemical similarities between the metabolism of early life and the redox couples in hydrothermal environments \cite{Martin2007}. \sed{There are also natural catalysts present in these systems that could have facilitated reactions now performed by enzymes \cite{Preiner2018}.} All these factors indicate that early life evolved in hydrothermal vent environments \cite{Martin2008}.  A key question then is what can reconcile the successful photochemical model of prebiotic chemistry with life's apparent roots in hydrothermal vent systems? 

One challenge faced by the hydrothermal origin scenario is how newly formed life could exploit the available energy gradients. Extant vent-dwelling life has developed complex molecular machines (Acetyl-CoA, ATPase, Carbon Monoxide Dehydrogenase, etc.) to take advantage of the rich resources hydrothermal vents offer, which it must construct and reproduce with fidelity. This indicates that protocellular life at hydrothermal vents must have come equipped with proteins, RNA and cell membranes, which in turn makes amino acids, nucleotides and vesicles minimal prerequisites for hydrothermal vent origin of life scenarios. However, at present, photochemically driven cyanosulfidic and cyanosulfitic protometabolism is the only path to forming all of these prerequisites in one environment, and this protometabolism proceeds only at surfaces exposed to sunlight. 

\sed{An alternative point of view is that \textit{both} the prebiotic chemistry and subsequent evolution occurred in underwater alkaline hydrothermal vents \cite{Russell2010,Harrison2018,Russell2018}}.   \rrsed{Recently, this hypothesis has gained support from the detection of nanomolar concentrations of amino acids of possible abiotic origin, found in the oceanic lithosphere proximal to underwater hydrothermal environments \cite{Menez2018}.} \sed{These claims are, however, disputed \cite{Jackson2016,Jackson2017a,Jackson2017b}. Most significantly, there is so far no experimental support for the generation of high} \rrsed{(>10\%)} \sed{and selective yields of amino acids and nucleotides from the molecular species expected to have been present in such an environment \cite{Cleaves2009,Danger2012} (although the question of what yields are necessary for the origin of life is also a matter of debate, e.g. \cite{Camprubi2017}).} 

The apparent exclusivity between the photochemical and vent-based scenarios reflects the two separate aspects of life’s origins they address: photochemistry successfully produces many of the building blocks for life, without, on its own, providing early life with a viable habitat; whereas vent-based scenarios succeed in linking early life to a viable habitat, but struggle to demonstrate how that life could have emerged from the ambient vent chemistry. In this context, surface hydrothermal vents provide a natural bridge between the surface scenario for prebiotic chemistry and the probable first habitat for early biology.  Such vents express a rich chemical diversity, with changes in composition and temperature with depth and time, which means a wealth of pre-biotic and early biotic environments may exist in superposition.

Surface hydrothermal vents may have been common on the subaerial mafic landmasses of the Early Earth \cite{nisbet2001_nature,mojzsis2001_nature,kamber2010_chemgeol}. Analogue systems on modern Iceland provide a natural laboratory for these early Earth conditions. Like their underwater cousins, the glaciovolcanic hydrothermal environments at Iceland's Kverkfj\"{o}ll volcano \sed{have diverse temperatures ranging from 0--95 $^{\circ}$C, and pH from 2--7.5. These pools are naturally anoxic,} and have \ce{H_2S} bubbling through them \cite{Cousins2013}. These surface vents will also be connected to deeper geothermal systems, which may produce fluids with elemental abundances fortuitously similar to those within the cell \cite{Mulk2012}. \sed{Many of the advantages for surface hydrothermal system overlap with the advantages of a sedimentary-hydrothermal environment, recently proposed as a context for life's origin by Westall \cite{Westall2018}.} Experiments in analogous geothermal pools have demonstrated that they can facilitate self-assembly \cite{Deamer2006}, and there is evidence that early life thrived in these environments \cite{Djokic2017}.  Surface hydrothermal vents therefore may provide a gradient along which proto-life can transition into life.

In considering the origin of life, we need to ask how different these hydrothermal systems may have been at the time when life originated versus today, and how these differences would have affected prebiotic chemistry. The photochemical prebiotic scenario has restrictive conditions on the environment in which it could take place:  In particular, the scenario requires high concentrations of HCN. It is very difficult to generate these high HCN concentrations, and may not be possible to sustain them, except in local environments where the carbon to oxygen ratio, C/O, is $\gtrsim 1$ \cite{Rimmer2018b}.

In this paper we demonstrate how ultra-reducing carbon- and nitrogen-rich magmas on the Early Earth can drive surface hydrothermal vents capable of supplying key feedstock molecules for prebiotic chemistry. We begin by establishing the scenario in which ultra-reducing carbon- nitrogen-rich magmas may have existed early in Earth's history, by reference to how Earth's atmosphere is constrained to have evolved (Section \ref{sec:bulk}).  We next investigate the gas-phase chemistry of magmas spanning a range of C/N/O (Section \ref{sec:variety}).  We then use the particular case of a carbon- and nitrogen-rich magma to model a simple surface hydrothermal vent scenario, and predict a suite of molecules that would be present therein (Section \ref{sec:vent}). Finally, we discuss the implications of these results and make some predictions for future geological measurements (Section \ref{sec:discussion}).

\section{Nitrogen and Carbon on Early Earth}
\label{sec:bulk}
A major factor in governing surface hydrothermal vent chemistry is the composition of the magma, which supplies the heat, juvenile volatile (C, H, N, O, S) and fluid soluble elements (e.g., Fe, and trace metals) to drive the system. On the modern Earth, magmatic gas chemistry varies significantly, with this diversity typically explained in terms of different outgassing pressure, temperature, and composition of the primordial melt (e.g., ref. \cite{Fischer2008}).  In this section we consider how magma volatile chemistry may have been different on the early Earth.

After its formation, the Earth would likely have inherited a primordial atmosphere via rapid outgassing and accretion of the surrounding gas \cite{Hayashi1979}.  The atmosphere at this stage was largely nebular, dominated by hydrogen and rich in ammonia, methane, carbon monoxide \cite{Zahnle2010}.  This primordial atmosphere would have transitioned via escape, rainout, and impact, into its secondary atmosphere.  Molecular nitrogen is thought to be the dominant constituent of this secondary atmosphere, followed by carbon dioxide and water \cite{Kasting1993}. 

The distribution of nitrogen among early Earth reservoirs is challenging to constrain, but we can get an indirect measure of the amount of nitrogen expected in the atmosphere from the total atmospheric pressure.  Observations of N and Ar isotopes constrain surface p\ce{N_2} to be less than $1.5\,$bar before $3\,$Ga \cite{Marty2013,avice2018_gca}. Although these papers permit an ancient surface pressure near that of the modern Earth, a simple fit to the data suggests total nitrogen partial pressures were likely somewhat lower, anywhere from 0 -- 1 bar (within $1\sigma$, see Appendix A). Recent models for nitrogen deposition and outgassing through the Earth's history also imply that the partial pressure of nitrogen was lower in the past \cite{Lammer2018}.

If the nitrogen partial pressure has been increasing over the first billion years of Earth history, then the nitrogen presently in our atmosphere must have been stored within the Earth's interior, and later outgassed.  This implies a nitrogen enrichment of the crust and/or upper mantle $\gtrsim 2$ Ga, creating the potential for magmas on the early Earth to have been enriched in either ammonia or \ce{N_2} depending on pressure and temperature \cite{Mikhail2014}.  

On the modern Earth, volcanic gases are particularly diverse, with measured eruptions exhibiting a range of C/O ratios from 0.003 (Usu, Japan) to 0.1 (Levotolo, Banda; \cite{Fischer2008}).  Whilst such carbon enrichments may come from remobilising crustal, biologically-derived, carbon \cite{mason2017_science}, there may have been alternative ways to achieve carbon-rich melts in a prebiotic context. 

Plausible constraints on the C/O ratios of magmas on the early Earth come from the oxidation state of the early, mafic, crust. \sed{The atmospheric composition of early Earth is not well-constrained before $\sim 3.5$ Ga, at which time it was probably weakly reducing \cite{Kasting1993}. Before this time,} \rrsed{it is thought that the atmosphere of the early Earth may have been dominated by \ce{CO_2} on the basis that 1) the inferred oxidation state of the mantle was close to the quartz-fayalite-magnetite mineral buffer (QFM, \cite{Trail2011}), and 2) that a \ce{CO2} dominated atmosphere is necessary for resolving the faint young sun paradox \cite{Shaw2008,Zahnle2010}} \rrsed{. However, cerium anomalies in Hadean zircons indicate that the crust-atmosphere system of the early Earth, older than 4$\,$Ga, was relatively more reducing than its contemporaneous mantle would suggest \cite{Yang2014,burnham2017_ngeo}. \citet{Yang2014} attribute this to impact bombardment `buffering' the $f_\mathrm{O_2}$ of the early crust.  Moreover, large amounts} of hydrogen from the primordial atmosphere could have been ingassed into the mantle and possibly the crust, creating the potential for ultra-reducing domains to form \cite{Sharp2017_chemgeol,Bali2013}. Together, these scenarios could locally lower the oxygen fugacity by 5 orders of magnitude below the IW buffer, allowing for regions where C/O > 1 and possibly C/H $\sim 1$ (if the hydrogen escapes the ultra-reducing domain, or reacts with the small amounts of available oxygen to form \ce{H_2O}). The effect of this excess carbon on the system's oxygen fugacity for a variety of C/O ratios is investigated in Appendix B.

We next investigate how the presence of nitrogen and/or carbon rich magmas would translate to the chemistry of early Earth volcanic gases, and thereby the production of key feedstock molecules for the photochemical origin of life scenario.

\section{The Gas-Phase Chemistry of Early Earth Basalts}
\label{sec:variety}

The possibility for enrichment of nitrogen and carbon in Early Earth magmas suggests four geochemical scenarios:
\begin{enumerate}
\item Nitrogen-Poor ($\ce{N}/\ce{O} \approx 7.7 \times 10^{-7}$) Carbon-Poor ($\ce{C}/\ce{O} \approx 0.5$, mean Oxidation State +3.8, $\Delta$NNO = -1)
\item Nitrogen-Rich ($\ce{N}/\ce{O} \approx 7.7 \times 10^{-4}$) Carbon-Poor (Section \ref{sec:noN-C})
\item Nitrogen-Poor, Carbon-Rich ($\ce{C}/\ce{O} \approx 3$, mean Oxidation State -0.3, $\Delta$NNO $\approx$ -11; Section \ref{sec:N-noC})
\item Nitrogen-Rich and Carbon-Rich (Section \ref{sec:N-C})
\end{enumerate}
See Appendix B for more details on the $f_{\rm O_2}$ each of these scenarios implies.

For each scenario we calculate the gas-phase chemistry.  Our aim is to identify whether any of the gas chemistries above produce significant quantities of cyanide and cyanoacetyline, the feedstock molecules for prebiotic photochemistry.  We perform these calculations by combining a model of the heterogeneous equilibria between melt and gas-phase for C/H/O-bearing species (D-Compress \cite{Burgisser2015}), with a thermochemical network to calculate the detailed molecular abundances in a C/H/O/N-bearing gas-phase (STAND2019 \cite{Rimmer2016,Rimmer2018b}). The network, STAND2019, has been benchmarked against a variety of planetary atmospheres, including the modern Earth, Mars, Jupiter, and hot and ultra-hot Jupiters, achieving results that agree with observations. In addition, its use of BURCAT and Benson values \cite{Burcat2005,Rimmer2016} to calculate the temperature-dependent Gibbs free energy for reversing reactions and achieving chemical equilibrium over reasonable time-scales finds agreement with combustion experiments \cite{Rimmer2016,Tsai2017}. This two-step approach is necessary to obtain mixing ratios for HCN and other hydrocarbon and nitrile species, which aren't included in most volcanic-gas codes, and indeed are not constituents in most modern (oxidised) volcanic gases.  

It is important to note that our approach does not produce self-consistent results, as we do not account for the feedback of the gas-phase chemistry derived from the thermochemical modeling onto the melt. However, this effect is likely to be small due to the dominant speciation of C and H in the melt as \ce{OH}, molecular \ce{H2O}, and carbonate ions, and the relative insolubility of reduced species such as \ce{H2} and \ce{CO} \citep{stolper1982_gca_b,dixon1988_epsl}. We run these calculations at \sed{a fixed fiducial} temperature \sed{of} 1200$^{\circ}$C, \sed{consistent with previous models \cite{Gaillard2014} and with what we undersand about the magma temperature during the Hadean \cite{Nisbet1993,Coogan2014}. We solve} for a closed system, where the gas travels with the melt as the pressure decreases, from 2 kbar to $\sim 1.5$ bar, with the initial conditions shown in Table \ref{tab:initial-conditions}. We modify the solver for the STAND network as follows.

\begin{table}
\centering
\begin{threeparttable}
\caption[Initial Conditions for D-Compress]{Initial Conditions for D-Compress \label{tab:initial-conditions}}
\begin{tabular}{lrr}
\toprule
Parameter & Value & Units\\
\midrule
Pressure & 2 & kbar\\
Temperature & 1473 & K\\
Background $\Delta$NNO & -1& \\
Fe$^{2+}$/Fe$^{3+}$ & 13.3& \\
Porosity Volume & 1.5398 & \% \\
Gas Weight & 0.5& \% \\
Gas Molar Weight & 52.52 & g$\,$mol$^{-1}$ \\
\ce{FeO} Weight & 10.24& \% \\
$\rho_{\rm melt}$ & 2.67 & g cm$^{-3}$\\ 
$\rho_{\rm gas}$ & 0.86 & g cm$^{-3}$\\
\ce{H_2O} melt weight & 0.098& \% \\
\ce{H_2} melt weight & 6.91 & ppb \\
\ce{SO_2} melt weight & 0.60& \% \\
\ce{H_2S} melt weight & 732 & ppm \\
\ce{CO_2} melt weight & 0.11& \% \\
\ce{S} melt weight & 0.36828& \% \\
\bottomrule
\end{tabular}
\end{threeparttable}
\end{table}

Instead of solving for the diffusion-photochemistry, we simply solve for:
\begin{equation}
\dfrac{dn_i(X)}{dt} = S_i(X) + P_i(X) - L_i(X)n_i(X),
\label{eqn:Magma-ARGO}
\end{equation}
where $S_i(X)$ [cm$^{-3}$ s$^{-1}$] is a source term capturing the transfer at the depth $i$ of species $X$ from the melt into the gas, $P_i(X)$ [cm$^{-3}$ s$^{-1}$] is the thermochemical production of $X$, $L_i(X)$ [s$^{-1}$] is the thermochemical loss of $X$, $n_i(X)$ [cm$^{-3}$] is the number density of $X$, and the the production and loss rates for $X$ are set such that the equilibrium abundance ($n_{\rm eq}$ [cm$^{-3}$]) is explicitly:
\begin{equation}
n_{\rm eq}(X) = \dfrac{P(X)}{L(X)}.
\end{equation}
We consider the melt to move at a range of rates from $v_{\rm melt} = 3 \times 10^{-5}$ m/s \cite{Turner2001} to 1 m/s \cite{mutch2018_ngeo}. The source term becomes:
\begin{equation}
S_i(X) = \dfrac{\big(n_{i+1}^0(X) - n_i^0(X)\big)v_{\rm melt}}{\Delta h},
\end{equation}
where $n_i^0$(X) [cm$^{-3}$] is the number density of species $X$ from D-Compress and $\Delta h$ [cm] is the change in crustal depth based on our step-size for the pressure. We only present the results for $v_{\rm melt} = 1$ m/s because the melt velocity has no impact on our results over the range we consider.

\begin{figure}
\centering
\includegraphics[width=0.75\linewidth]{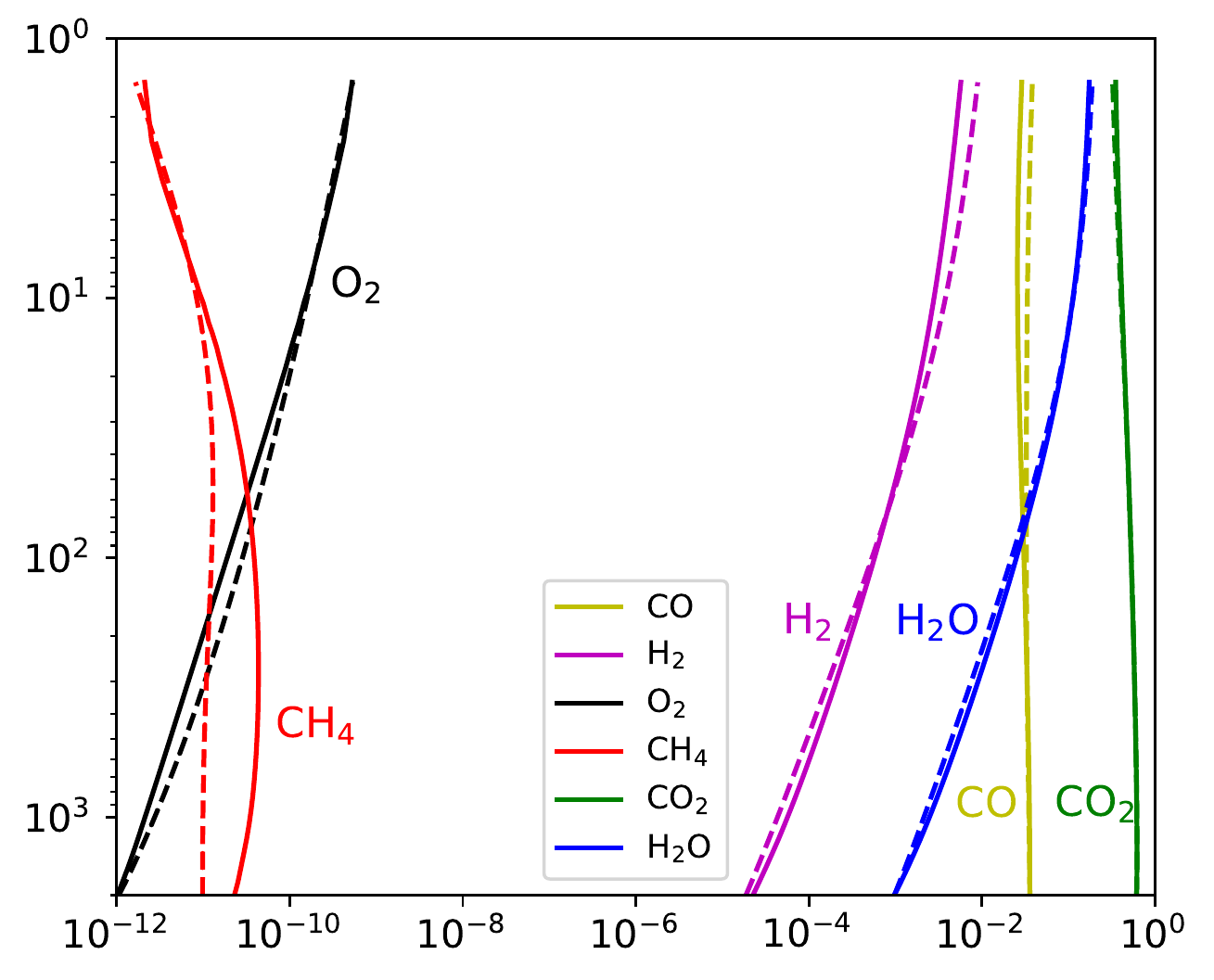}
\caption{{\bf A comparison of the mixing ratios (x-axis) as a function of pressure (bar, y-axis) between STAND and a modified version of ARGO (solid) with the D-Compress gas-phase (dashed).} \label{fig:compare}}
\end{figure}

We present a comparison of the results for the nitrogen-poor carbon-poor case between D-Compress and STAND (Fig. \ref{fig:compare}). The results compare well with all species except methane, for which there is an order of magnitude discrepancy at greater than 10 bar, although both nonetheless predict trace ($< 0.1$ ppb) quantities. 

\begin{figure}
\centering
\includegraphics[width=0.5\linewidth]{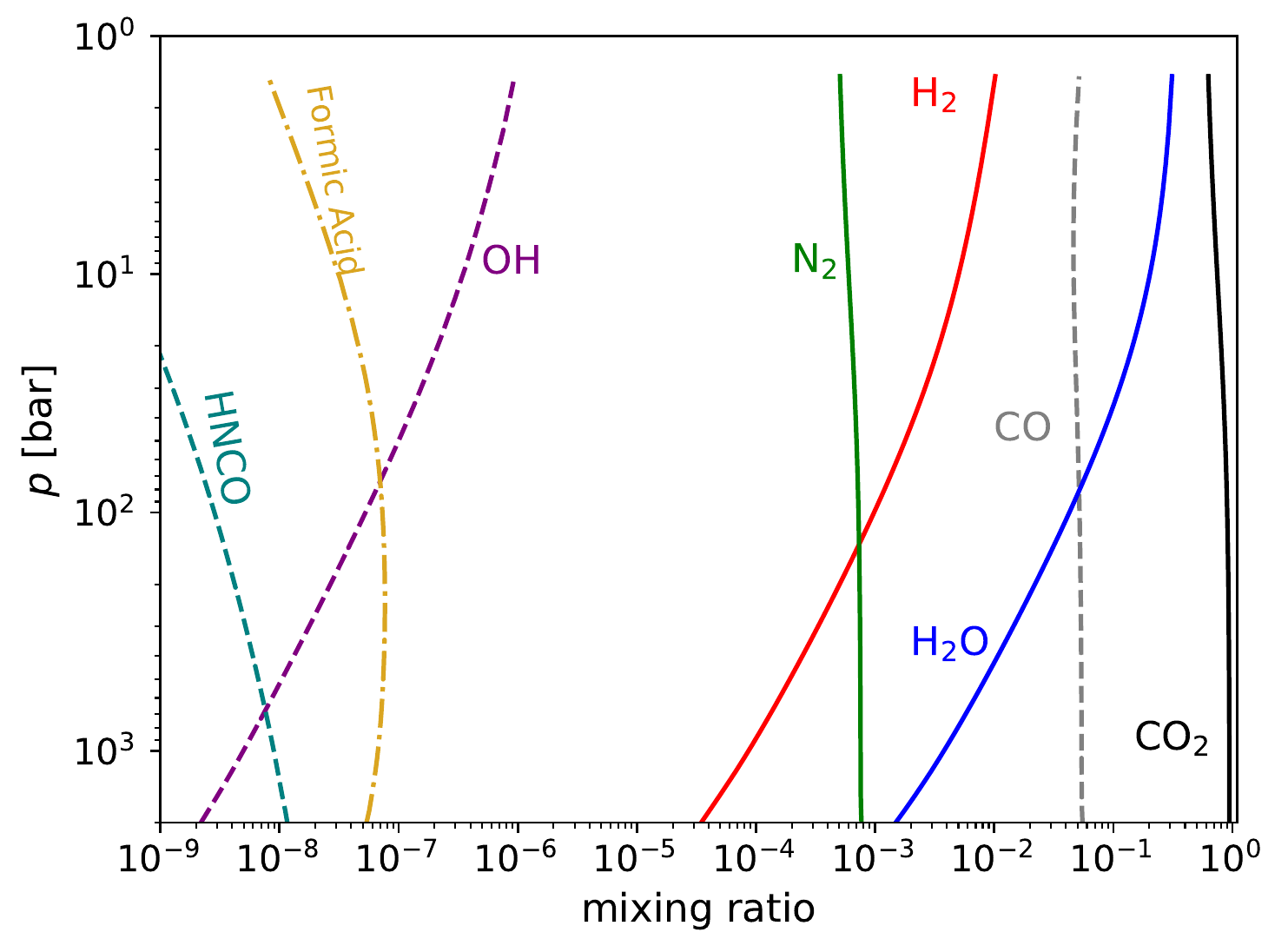}
\includegraphics[width=0.5\linewidth]{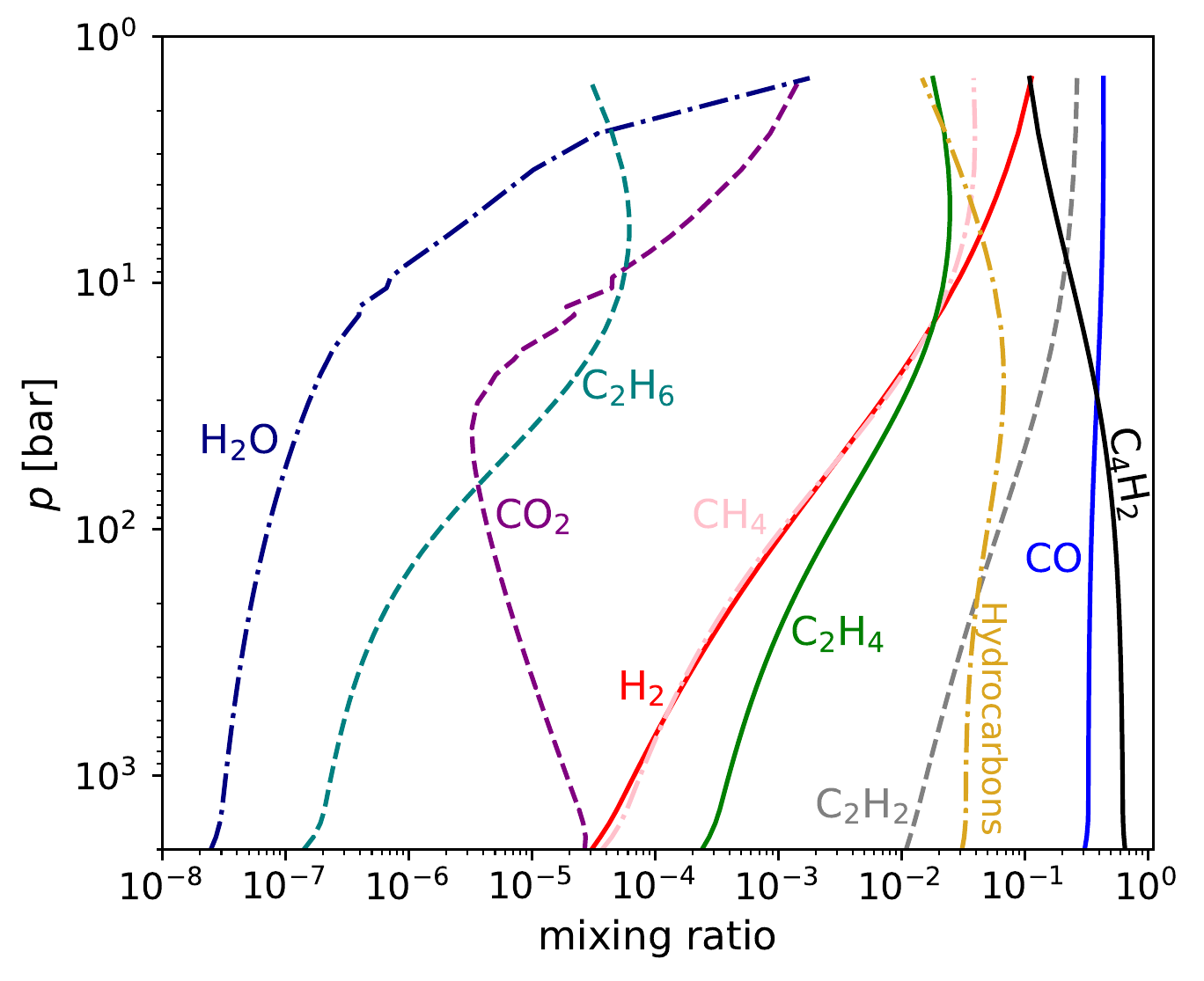}\\
\includegraphics[width=0.5\linewidth]{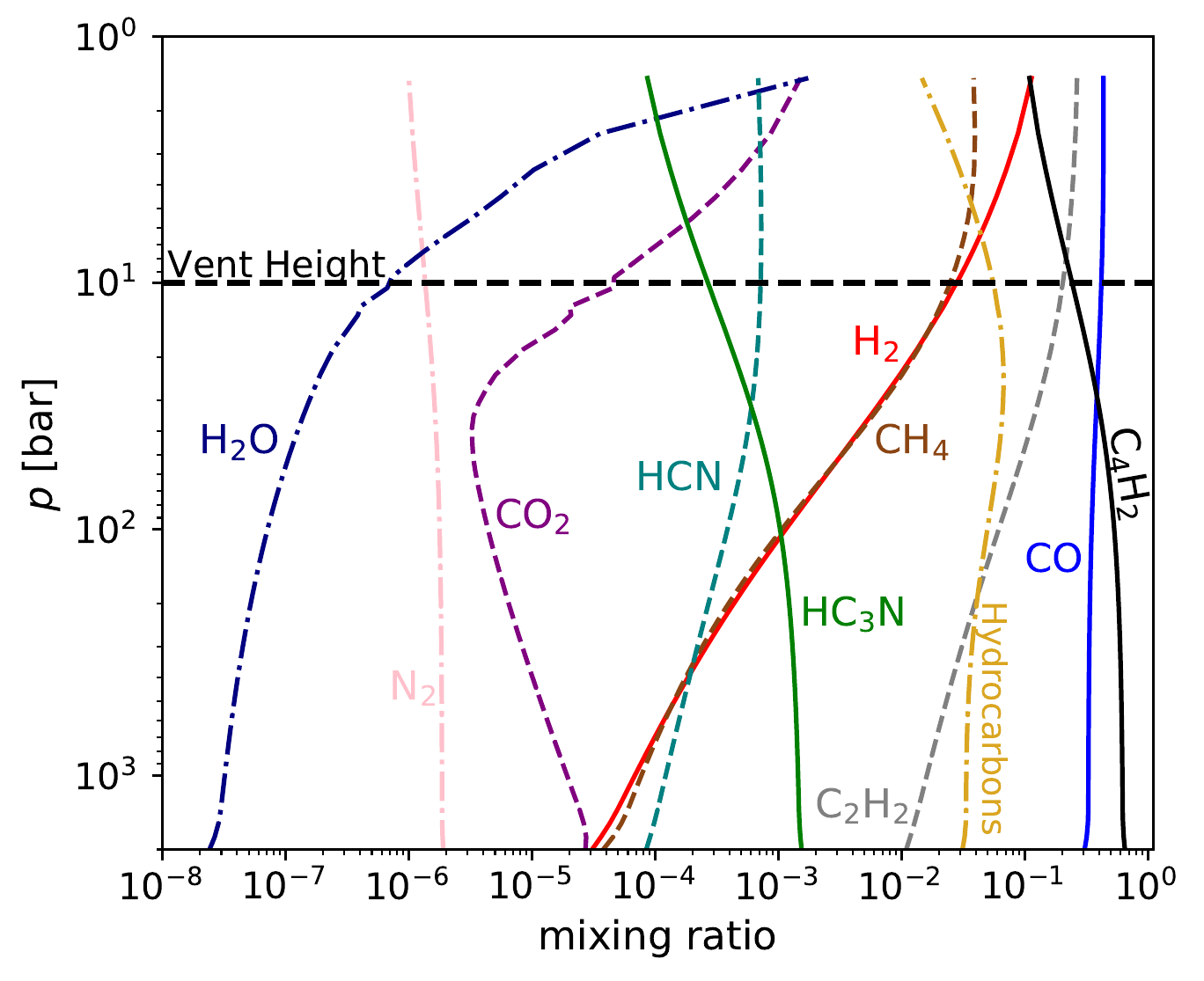}
\caption{Mixing ratios of species in melt+gas-phase system at 1200 $^{\circ}$C versus pressure, $p$ [bar], for three different scenarios (see Section \ref{sec:variety} for details), carbon-poor and nitrogen-rich (top), carbon-rich nitrogen-poor (middle), and carbon-rich nitrogen-rich (bottom). \label{fig:magma}}
\end{figure}

The results for the next three cases are shown in Figure \ref{fig:magma}.

\subsection{Nitrogen-Rich Carbon-Poor Magma}
\label{sec:N-noC}

In the nitrogen-rich carbon-poor scenario, we increase the gas-phase nitrogen from 1 ppm to 0.1\%, in the form of \ce{N_2} at the 2 kbar level. This is the most stable form for the nitrogen present at this temperature and pressure \cite{Mikhail2014}. The gas-phase chemistry is very close to the nitrogen-poor carbon-poor scenario, except for the presence of \ce{N_2} at 0.1\%. The \ce{N_2} is relatively unreactive, as is the \ce{CO_2} and \ce{CO}. Changes with pressure are seen for \ce{H_2O}, \ce{H_2} and the sulfur-bearing species, with most of these changes driven by loss of these species from the melt to the gas phase.

\subsection{Nitrogen-Poor and Carbon-Rich Magma}
\label{sec:noN-C}

Increasing the concentration of carbon has a major impact on gas-phase chemistry.  We introduce the carbon in the form of \ce{C_2H}, a thermochemically stable form of carbon at high temperatures and pressures, depending on the amount of available hydrogen \cite{Hu2014}. At higher pressures, our model predicts that virtually all of the injected \ce{C_2H} is converted to diacetylene (\ce{C_4H_2}). Throughout this and the next couple sections, we write reactions with arrows in both directions, \ce{<<=>} and \ce{<=>>}. As the chemistry has effectively reached equilibrium, these arrows note in which direction the equilibrium tends when the pressure decreases, whether the equilibrium goes in the direction \sed{from product to reactant (\ce{<<=>}) or from reactant to product (\ce{<=>>}).}

Effectively all of the \ce{CO_2} is released at the 2 kbar level, and it reacts away to form CO, CH$_4$, and C$_2$H$_2$ via the chemical sequence (for $p > 100$ bar; \ce{M} represents any third body present in the gas-phase):
\begin{align}
\ce{C_4H_2} + \ce{M} &\ce{<<=>} 2\ce{C_2H} + \ce{M}\notag\\
\ce{H_2} + \ce{C_2H} &\ce{<=>>}\ce{C_2H_2} + \ce{H}\notag\\
\ce{CO_2} + \ce{H} &\ce{<=>>} \ce{CO} + \ce{OH}\notag\\
\ce{C_2H_2} + \ce{OH} &\ce{<=>} \ce{CH_3} + \ce{CO}\notag\\
\ce{C_4H_2} + \ce{CH_3} &\ce{<<=>} \ce{C_4H} + \ce{CH_4}\notag\\
\ce{C_4H} + \ce{H_2} &\ce{<=>>} \ce{C_4H_2} + \ce{H}\notag\\
\ce{C_2H} + \ce{H} + \ce{M} &\ce{<=>>} \ce{C_2H_2} + \ce{M}\notag\\
\cmidrule(lr){1-2}
\ce{CO_2} + \ce{C_4H_2} + 2 \ce{H_2} &\ce{<=>>} 2\ce{CO} + \ce{CH_4} + \ce{C_2H_2}.
\end{align}
This conversion is most limited by the reactions involving H and \ce{H_2}, and the main cause of this limit is the availability of hydrogen. Note that, in the carbon-rich cases, there is very little gas-phase water. This is because the water is destroyed, providing the hydrogen, and more hydroxyl radicals, in the following way:
\begin{align}
\ce{C_4H_2} + \ce{M} &\ce{<<=>} 2\ce{C_2H} + \ce{M} \notag\\
\ce{C_2H} + \ce{H_2O} &\ce{<=>>} \ce{C_2H_2} + \ce{OH}\notag\\
\ce{C_2H} + \ce{H_2O} &\ce{<=>>} \ce{HCCO} + \ce{H_2} \notag\\
\ce{HCCO} + \ce{C_2H_2} &\ce{<=>>} \ce{C_3H_3} + \ce{CO} \notag\\
\cmidrule(lr){1-2}
\ce{C_4H_2} + 2\ce{H_2O} &\ce{<=>>} \ce{C_3H_3} + \ce{CO} + \ce{OH} + \ce{H_2}. \label{eqn:co2}
\end{align}
This produces both hydroxyl radicals, carbon monoxide and what is effectively ``soot''\footnote{The model is only accurate for a handful of species with more than two carbon atoms, and quite a bit of the carbon-rich chemistry artificially terminates in species like \ce{C_3H_3}. In reality, the polymerization chemistry would continue limited only by molecular size, available material and time, into very large clumpy soots.}, as well as the molecular hydrogen. The atomic hydrogen arises from the molecular hydrogen from a great variety of reactions.

The rapid upswing of both the carbon dioxide and molecular hydrogen is due to the release of large quantities of water at lower pressures, and the consequent reaction:
\begin{equation}
\ce{CO} + \ce{H_2O} \rightleftharpoons \ce{CO_2} + \ce{H_2}.
\end{equation}
The increase in the amount of water also drives up the OH abundance which, along with the increased amount of \ce{H_2} and \ce{CO_2}, more rapidly converts diacetylene (\ce{C_4H_2}) to acetylene (\ce{C_2H_2}) via Reaction (\ref{eqn:co2}).

These series of reactions also show how \ce{C_2H2}, methane (\ce{CH_4}) and carbon monoxide (\ce{CO}) are generated. Within most of this temperature-pressure regime ($p > 100$ bar), with sufficient hydrogen, methane should be more abundant than CO, but the methane formation in this case is severely hydrogen-starved.

\subsection{Carbon-Rich Nitrogen-Rich Magma}
\label{sec:N-C}

When both carbon and nitrogen are abundant, the chemistry becomes far more complex. We can nevertheless describe the dominant reactions that destroy some triple bonds and make others. It is possible in this scenario to break the typically sturdy triple bond between the atoms of \ce{N\equiv N}, often ultimately bringing about another triple bond, \ce{C\equiv N}. The dominant mechanism by which this conversion occurs is:
\begin{align}
\frac{1}{2}\big(\ce{C_4H_2} + \ce{M} &\ce{<<=>} 2 \ce{C_2H} + \ce{M}\big)\notag\\
\ce{N_2} + \ce{H} + \ce{M} &\ce{<=>>} \ce{N_2H} + \ce{M}\notag\\
\ce{N_2H} + \ce{CO_2} &\ce{<=>>} \ce{NO} + \ce{HNCO}\notag\\
\ce{NO} + \ce{C_2H_2} &\ce{<=>>} \ce{HCN} + \ce{CO} + \ce{H}\notag\\
\ce{HNCO} + \ce{H} &\ce{<=>>} \ce{HNC} + \ce{OH}\notag\\
\ce{HNC} + \ce{M} &\ce{<=>>} \ce{HCN} + \ce{M}\notag\\
\ce{HCN} + \ce{C_2H} &\ce{<=>>} \ce{HC_3N} + \ce{H}\notag\\
\cmidrule(lr){1-2}
\ce{N_2} + \ce{CO_2} + \frac{1}{2}\ce{C_4H_2} + \ce{C_2H_2} &\ce{<=>>} \ce{HC_3N} + \ce{HCN} + \ce{CO} + \ce{OH}. \label{eqn:cyanoacetylene}
\end{align}
This is how, at high pressures, diacetylene (\ce{C_4H_2}) reacts with the product of molecular nitrogen (isocyanic acid, \ce{HNCO}) to form hydrogen cyanide (\ce{HCN}) and cyanoacetylene (\ce{HC_3N}). At high pressures, more cyanoacetylene is produced than hydrogen cyanide, because of the last reaction in the above sequence:
\begin{equation}
\ce{HCN} + \ce{C_2H} \ce{<=>>} \ce{HC_3N} + \ce{H},
\end{equation}
which will consume a good fraction of the hydrogen cyanide produced by Reaction (\ref{eqn:cyanoacetylene}).

At lower pressures, however, where there is more hydrogen and acetylene, and the equilibrium balance also shifts to favor hydrogen cyanide:
\begin{align}
\ce{HC_3N} + \ce{H} &\rightleftharpoons \ce{CN} + \ce{C_2H_2}\notag\\
\ce{HC_3N} + \ce{H} &\rightleftharpoons \ce{HCN} + \ce{C_2H}\notag\\
\ce{CN} + \ce{C_2H_2} &\ce{<=>>} \ce{HCN} + \ce{C_2H}\notag\\
2(\ce{C_2H} + \ce{H_2} &\ce{<=>>} \ce{C_2H_2} + \ce{H})\notag\\
\cmidrule(lr){1-2}
2(\ce{HC_3N} + \ce{H_2} &\ce{<=>>} \ce{HCN} + \ce{C_2H_2}). \label{eqn:cyanide}
\end{align}
By these Reactions (\ref{eqn:cyanoacetylene})-(\ref{eqn:cyanide}), cyanoacetylene and cyanide overwhelm molecular nitrogen and trade places with each other as the dominant nitrogen-bearing species in the volcanic gas-phase.

\section{Surface Hydrothermal Vents with Carbon- Nitrogen-Rich Magma Degassing}
\label{sec:vent}

\begin{figure}
\centering
\includegraphics[width=0.7\linewidth]{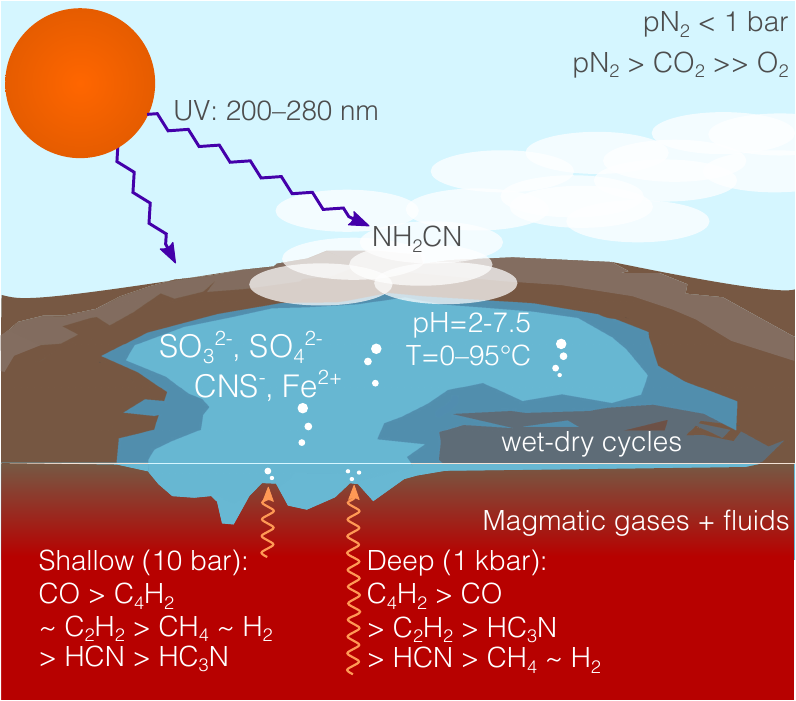}
\caption{A surface hydrothermal vent fed by a carbon-rich nitrogen-rich magma. The edges of the pool may expand and recede, providing dry-wet cycles. The water of the pool would plausibly be enriched by sulfites (\ce{SO_3^{2-}}), sulfates (\ce{SO_4^{2-}}), and Iron(II). Various amounts of diacetylene (\ce{C_4H_2}), acetylene (\ce{C_2H_2}), cyanoacetylene (\ce{HC_3N}), hydrogen cyanide (\ce{HCN}), hydrogen (\ce{H_2}) and methane (\ce{CH_4}) bubble up out of the pool in ratios determined by the depth at which the gas is quenched. One released into the air above the pool, these species are photolysed and react to produce small amounts of cyanamide. The cyanide in the pool will react with the sulfates to slowly form thiocyanate (\ce{CNS^-}), and will react with solvated electrons to produce simple sugars and amino-acid precursors.  \label{fig:schematic}}
\end{figure}

The previous section demonstrates the range of magmatic gas chemistries that could be feeding surface hydrothermal vents on the early Earth.  However, only a subset of these outcomes provide prebiotically useful chemistries: the UV-photochemical prebiotic scenario has specific requirements for its feedstock molecules, requiring cyanide to initiate and proceeding from this to make use of cyanoacetyline. The nitrogen-poor magmas do not have enough nitrogen to build up high concentrations of cyanide or cyanoacetylene, and the most stable form for nitrogen in the nitrogen-rich carbon-poor magmas is \ce{N_2}. In contrast, both cyanide and cyanoacetyline are orders of magnitude more abundant during degassing of ultra-reduced carbon- nitrogen-rich magmas than in the other scenarios. We therefore proceed with these gases as feeding a surface hydrothermal vent possessing sulfite-, sulfate-, and \ce{Fe^{2+}}-rich water, as shown in Figure \ref{fig:schematic}.

We can estimate the concentrations of key prebiotic molecules in water interacting with gas using Henry's Law, under the assumption that the gas quickly cools in contact with the water to near room temperature. A range of pressures could be chosen at which magmatic gases equilibrate with the water, and higher pressures will result in a more cyanoacetylene-rich gas, lower pressures in a more cyanide-rich gas. The surface hydrothermal vents at Kverkfj\"{o}ll outgas \ce{H_2S} \cite{Cousins2013}, suggesting a quenching of gas-phase chemistry at a pressure of $\sim 10$ bar, and hence we select this pressure when calculating gas-water equilibrium in this section.

The Henry's Law constants we use are mostly taken from Sander et al. (2015) \cite{Sander2015}. For those constants that are not included in the literature, we use the estimates from ChemAxon\footnote{\url{http://www.chemaxon.com}}. \sed{Table \ref{tab:sol} gives concentrations from the gas impinging on water at 10 bar. It may be that some of the enriched water will concentrate via evaporation at the shores of the pools at the top of the vent, and so the concentrations of some of the tabulated species may be enhanced.}  The concentrations for hydrogen sulfide, sulfur dioxide, acetylene, diacetylene and cyanoacetylene are \sed{near saturation}.

The sulfur dioxide absorbed into the vent water will quickly be dissociated, ($pK_a (\ce{SO_2}) = 1.8$), reacting with water to form \cite{Ranjan2018}:
\begin{equation}
\ce{SO_2} + 2\ce{H_2O} \rightleftharpoons \ce{HSO_3^-} + \ce{H_3O^+},
\end{equation}
which can be reduced further to form sulfite (\ce{SO_3^{2-}}) and sulfate (\ce{SO_4^{2-}}), or oxidized to form sulfuric acid (\ce{H_2SO_4}) \cite{Ranjan2018}. Because there is a relatively small amount of available oxygen, the bulk of the sulfur will be in the form of bisulfite (\ce{HSO_3^-}) and hydrogen sulfide (\ce{HS^-}); the relatively small amount of sulfate will react rapidly with the hydrogen cyanide to form thiocyanate (\ce{CNS^-}). We propose that the thiocyanate observed in Yellowstone springs \cite{Kamyshny2014} is a sign that this same chemistry occurs, at much smaller concentrations and in a much more oxidized local environment, in volcanic environments today.

We expect for the Carbon-poor Nitrogen-poor case that more than 90\% of the iron that would be present in the vent environment is in the form of \ce{Fe^{2+}}. Most of the iron in this case would be expected to be in the form of FeO as silicates, which is effectively insoluble, but with significant fractions of iron in solution in the form of Iron (II) Sulfate and Iron (II) Nitrate. The Iron (II) Sulfate would saturate in the intersecting water at 2 M concentrations, based on its measured solubility. Iron (II) Nitrate would saturate in the intersecting water at 3 M concentrations, based on calculated solubility\footnote{a theoretical $\log S = 0.49$ from ChemAxon \url{(http://www.chemaxon.com)}}. For the ultra-reducing cases, it is plausible that the iron may be even further reduced before its release into the vent water.

\begin{table}
\centering
\caption{Concentrations of species in water that intersects the C-rich/N-rich flow at 10 bar \label{tab:sol}}
\begin{tabular}{lclcl}
\hline
Species & \multicolumn{2}{c}{$p$} & \multicolumn{2}{c}{$c$} \\
\hline\hline
Hydrogen Sulfide (\ce{H_2S}) & 0.2 & bar & 20 & mM\\
Sulfur Dioxide (\ce{SO_2}) & 3 & bar & 3 & M \\
Carbon Monoxide (\ce{CO}) & 3.8 & bar & 3.8 & mM \\
Methane (\ce{CH_4}) & 63 & mbar & 88 & $\mu$M \\
Acetylene (\ce{C_2H_2}) & 1.8 & bar & 46 & mM \\
Diacetylene (\ce{C_4H_2}) & 4.2 & bar & 0.19 & M \\
Hydrogen Cyanide (HCN) & 7.5 & mbar & 90 & mM \\
Cyanoacetylene (\ce{HC_3N}) & 5.6 & mbar & 0.23 & M \\
Acrylonitrile (\ce{C_3H_3N}) & 0.28 & mbar & 3.2 & mM \\
Formaldehyde (\ce{HCHO}) & 0.29 & mbar & 1.1 & mM \\
\hline
\end{tabular}
\end{table}

Some cyanamide can also generated photochemically near the interface of the atmosphere and the vent surface (discussed by \cite{Rimmer2018b}):
\begin{align}
\ce{HCN} + \ce{O} &\rightarrow \ce{NCO} + \ce{H}, \notag\\
&\rightarrow \ce{CO} + \ce{NH}, \notag\\
\ce{NCO} + \ce{H} &\rightarrow \ce{CO} + \ce{NH}, \notag\\
\ce{NH} + \ce{CH_4} + \ce{M} &\rightarrow \ce{NH_2CN} + \ce{M}.
\end{align}
But this photochemically generated cyanamide only achieves mixing ratios of at most $10$ ppb, which can be enhanced through absorption into rain droplets and subsequent evaporation, even then only reaching concentrations $< 10$ $\mu$M. 

\section{Discussion}
\label{sec:discussion}

In this paper we have shown how ultra-reducing carbon- and nitrogen-rich surface hydrothermal vents are an environment that can bridge two of the main origin of life scenarios: marine hydrothermal vents and surface UV-photochemistry.  Surface hydrothermal vents achieve this by providing the favorable habitat early life appears to have occupied, whilst also providing the feedstock molecules and access to stellar light needed for UV-photochemistry to build life's building blocks.  
We first considered how carbon- and nitrogen-rich reservoirs may have existed in the early earth, based observationally on the probable low p\ce{N_2} of the early Earth atmosphere. Then we calculated the gas-phase chemistry of four different magma compositions (standard oxidising, carbon-rich, nitrogen-rich, and carbon- and nitrogen-rich).  Only the ultra-reduced and carbon- and nitrogren-rich case produced substantial cyanide and cyanoacetyline, key feedstock molecules for prebiotic chemistry, so we focused on modelling this gas chemistry's interaction with a surface hydrothermal vent system. 

From an origin of life perspective, our key result is that a surface hydrothermal pool fed by such gases would be enriched, and in some cases saturated, with acetylene, diacetylene, hydrogen cyanide, cyanoacetylene, acrylonitrile, sulfur dioxide, hydrogen sulfide, and \ce{Fe^{2+}}, providing a veritable ``buffet lunch'' for subsequent prebiotic chemistry. The presence of \ce{Fe^{2+}} in a cyanide-rich environment opens a way to connect this scenario to closely related scenarios involving organometallic chemistry \cite{Ritson2018}. Each species' concentration was considered independently, and it is important to note that the absorption efficiency of some of these chemical species into the water may be changed by the presence of other chemical species, which could enhance or interfere with the absorption efficiency and resulting concentrations. Traces of cyanamide, significant for the atmospheric chemist but not for the prebiotic chemist, are predicted to be produced from photolysis of hydrogen cyanide and subsequent reaction with molecular hydrogen. 

\subsection{Challenges of making life on land}
\sed{The choice of surface hydrothermal vents as a context for life's origin does, however, present challenges not faced by their deep marine counterparts. One of the most important of these challenges lies with the very resource making surface environments appealing in the first place: UV light.   UV-driven chemistry is constructive for synthesizing and activating nucleotides and amino acids (e.g. \cite{Mariani2018}), but is destructive for more complex organic systems, such as, for example, DNA strands \cite{Matsunaga1991,Karentz1991}. UV damage of strands of organic molecules is problematic for life and proto-life that is less than tens of meters under pure water \cite{Quickenden1980}, but the attenuation depth for metal-rich vent water is unknown. Extant natural bodies of fresh water exhibit UV attenuation on the order of a meter or less \cite{Laurion2000}. How restrictive this is for an origin of life in shallower water bodies is unclear though, when shelter may be readily provided under outcrops, vent cavities, or vesicles in the rock.}

\sed{The surface of the early Earth would also have been subjected to intense meteoritic and cometary bombardment.  Impactors avoiding breakup in the atmosphere would sterilize the surface region near where they struck \cite{Abramov2009}. If the Earth experienced a spike in the impact rate at $\sim 3.9$ Ga, a Late Heavy Bombardment (LHB), this would be especially deleterious for surface life \cite{Abramov2009,Abramov2013}. In this event, the immediate surface would probably be uninhabitable, but a large fraction of the near-surface and subsurface could still support microbial life \cite{Abramov2009}. Any initially-subaerael scenario could have taken place sufficiently before the LHB in order to allow the resulting life or proto-life to descend into the near-surface or sub-surface, either if the environment itself falls into the ocean, or if the life disperses into the subsurface. Alternatively, impact-induced hydrothermal systems may have provided the ideal surface environemnts for life's origin after the LHB \cite{Abramov2009}.}

\sed{The Late Heavy Bombardment only poses a problem for surface hydrothermal vents if it occurred. The evidence for the LHB can be explained either by a massive spike in the impact rate at $\sim 3.9$ Ga or by a gradual, monotonic decrease in the impact rate from accretion until $\lesssim 3$ Ga \cite{Chapman2007,Boehnke2016}. If there was no LHB, then the challenges for the survival of life near the immediate surface, sufficiently after accretion, is far less severe. }

\subsection{Making the surface hydrothermal vent origin of life scenario testable}
We have focused on a local hydrothermal environment, and by all indications local environments are ``big enough'' for the origin of life. Local environments are not generally accessible within the geological record, however, and this is especially the case within the Early Archean and Hadean for which the geological record reduces to refractory mineral phases such as zircon.  \sed{As with many, if not all, origins of life scenarios, this deficit of geological record over the key time period in Earth history poses a difficulty for validating the surface hydrothermal vent scenario.}  

There are, though, two approaches we can take at present to elevate our proposed scenario from a ``just-so'' story into a testable hypothesis: \sed{first,} to draw connections between the local environment and the global environment on one hand, and the local environment and the lab on the other.  For example, one important way in which zircon-based inferences of early Earth geology do support the viability of our model is that they indicate that subaerial landmasses were present and interacting with the hydrosphere from 4 Ga \cite[e.g.,][]{mojzsis2001_nature}. 

\sed{It will be difficult to perform geochemical and prebiotic experiments suggested by this scenario: experiments need to investigate whether all of these chemicals, some very reducing, some oxidizing, can coexist together in a hydrothermal environment, and could survive the transition from the 1200$^{\circ}$ C magma to temperate waters} and, if so, over what timescales. \sed{A more promising approach will be to perform prebiotic experiments using the output from our model or experiments. In this way, we may discover new unexpected} species generated within a simulated nitrogen-rich ultra-reducing exposed vent, to see if these species disrupt or enhance the subsequent chemistry. In this way, we can turn the concept of prebiotic plausibility on its head, and use prebiotic chemistry to inform early Earth geochemistry. It is our opinion that both prebiotic chemistry and geochemistry can inform each other, both about the origin of life and, given life's emergence, about what local environments were plausibly present on the early Earth.

As emphasised, highly reducing environments, at least locally, are key for this scenario's successful production of prebiotic feedstock.  There are plausible endogenic (hydrogen fluxes from serpentinisation...) and exogenic (reduced impactors...) sources of reducing power on the early Earth.  Constraining the ubiquity of such environments requires further work to determine interactions at the interfaces of the hydrosphere with the ultra-mafic crust and the chemistry of late accretion during the epoch of life's emergence. Models of how local environments arise from global environments (e.g., how the formation of subaerial hydrothermal systems are linked to mantle temperature and ocean water mass), combined with new evidence of the nature of the global environment, will give us insight into the probability of local environments having existed. At the same time, laboratory simulations of these local environments will provide a way to test whether our predictions about the specific nature of these environments are accurate. Both of these avenues are fields for future exploration as researchers continue to investigate the unsolved problem of life's origin on Earth.


\funding{This research was funded by the Simons Foundation (SCOL awards 599634).}

\acknowledgments{We are grateful for fruitful discussions about this work that took place at the Lorentz Center Workshop ``Roadmap for Universal Life''. PBR would additionally like to thank Claire Cousins for helpful discussions about exposed hydrothermal vents.}

\conflictsofinterest{The authors declare no conflict of interest.} 



\appendixtitles{yes} 
\appendixsections{multiple} 
\appendix
\section{Partial Pressure of Nitrogen in Early Earth Atmosphere}
\begin{figure}[ht!]
\centering
\includegraphics[width=0.85\textwidth]{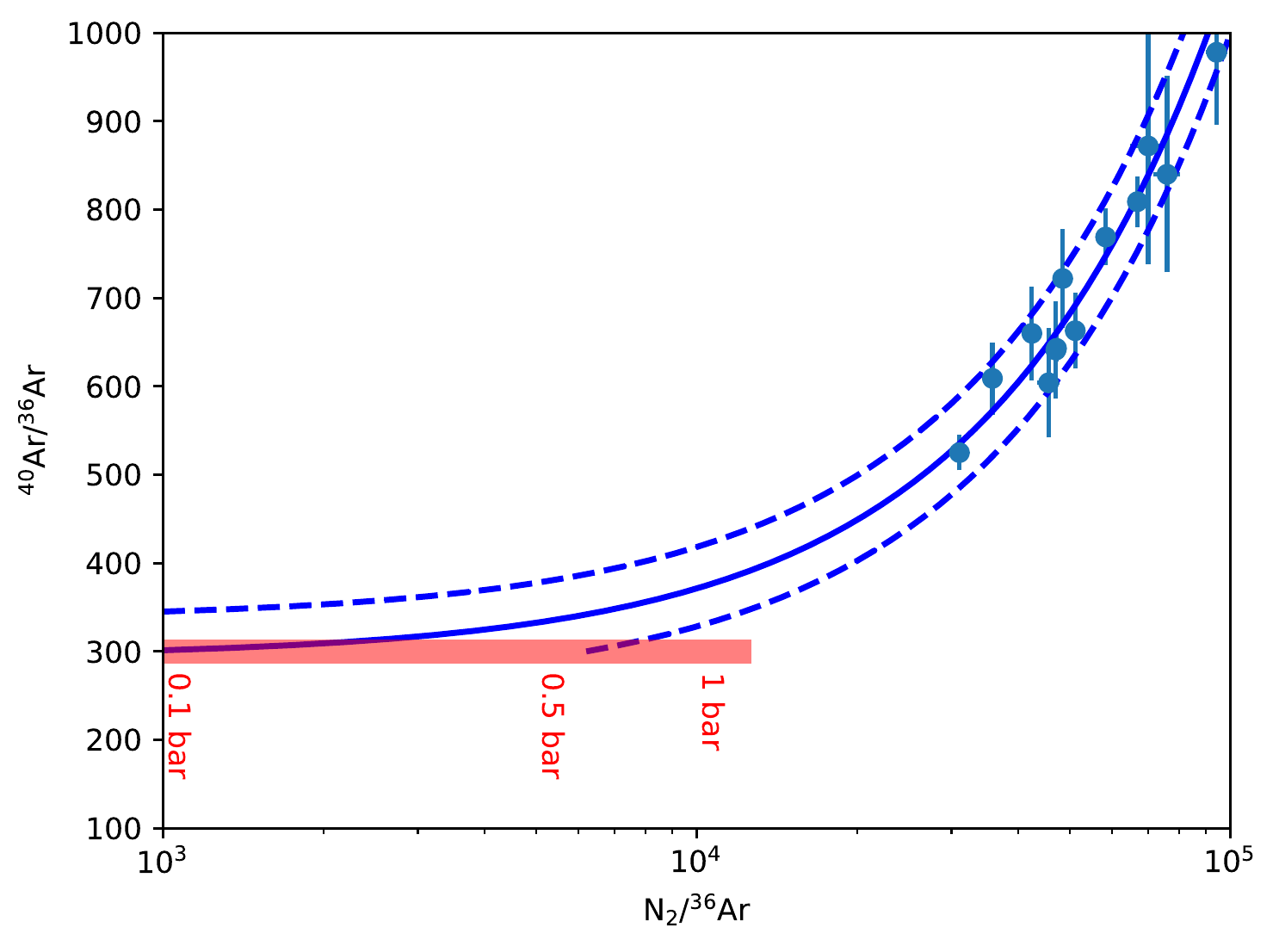}
\caption{The ratio \ce{N_2}/\ce{^{36}Ar} as a function of  \ce{^{40}Ar}/\ce{^{36}Ar} with a linear fit to the data$^{X}$ \label{fig:pn2}}
\end{figure}

We take the data from Avice et al (2018) \cite{avice2018_gca}, their Table 2, and reproduce the correlation found in their Figure 4(a). We perform a linear regression on the data and find a fit of:
\begin{equation}
\dfrac{\ce{N_2}}{\ce{^{36}Ar}} = \big(128.6 \pm 5.5 \big) \Bigg(\dfrac{\ce{^{40}Ar}}{\ce{^{36}Ar}}\Bigg)
- \big(37740 \pm 3740\big).
\end{equation}
The fit is shown as a solid blue line in Fig. \ref{fig:pn2}, where the dotted blue lines show the $1\sigma$ standard deviation from the fit, and the red line at \ce{^{40}Ar}/\ce{^{36}Ar} = 300 represents the inferred range of $p(\ce{N_2})$ from 0 - 1 bar. The data justify inferring any value of $p(\ce{N_2})$ of 0 - 1 bar at more than 3 Gya, to within $2\sigma$.

\section{Oxygen Fugacity in the Carbon-Rich Magma}

\begin{figure}
\centering
\includegraphics[width=\textwidth]{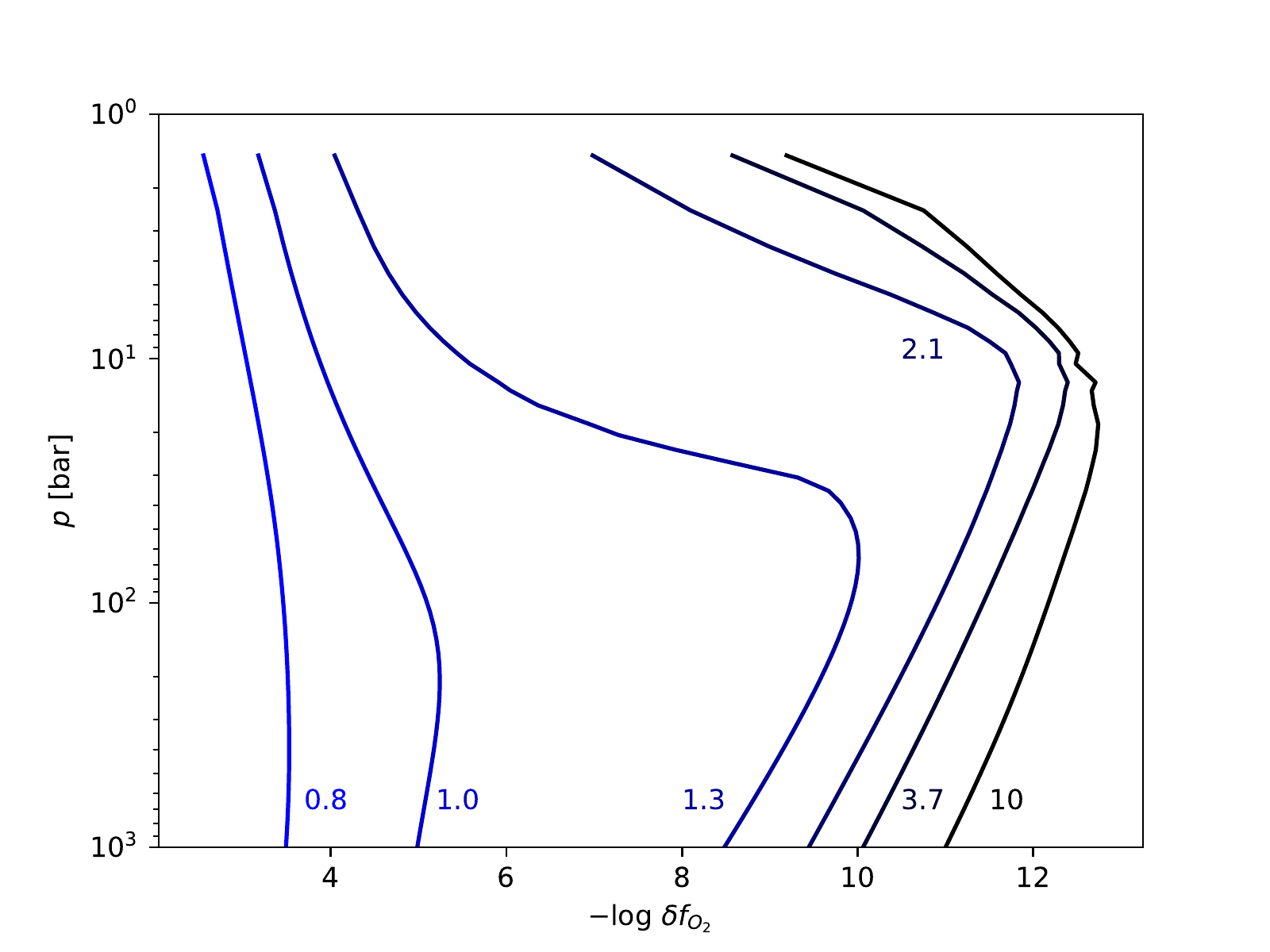}
\caption{The change in Oxygen fugacity, in the form $-\log \, \delta f_{\rm O_2}$, as a function of pressure for various C/O ratios from 0.8 (blue) -- 10 (black), with the values noted next to the relevant curves. \label{fig:fo2}}
\end{figure}

\begin{figure}
\centering
\includegraphics[width=0.65\textwidth]{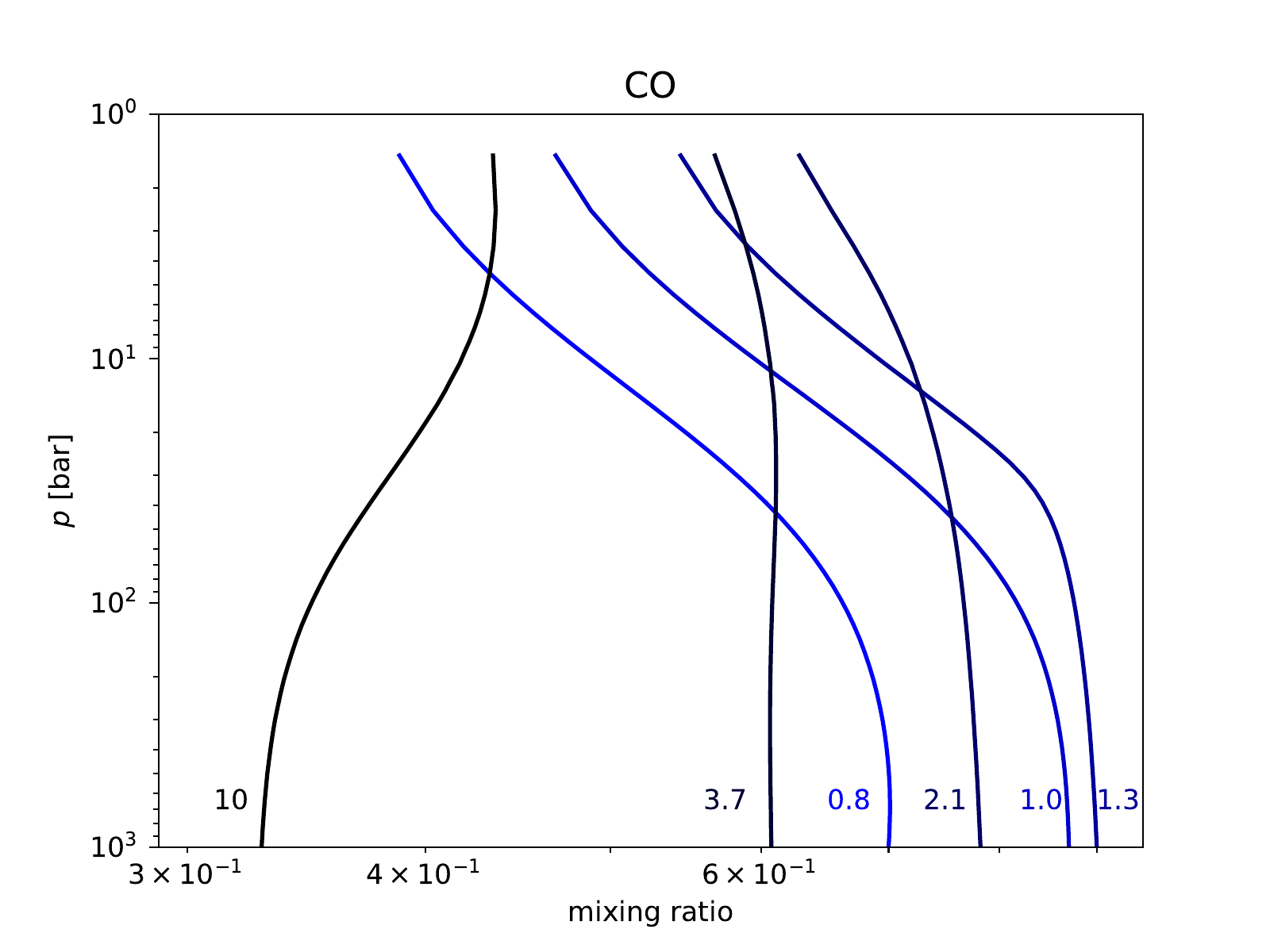}
\includegraphics[width=0.65\textwidth]{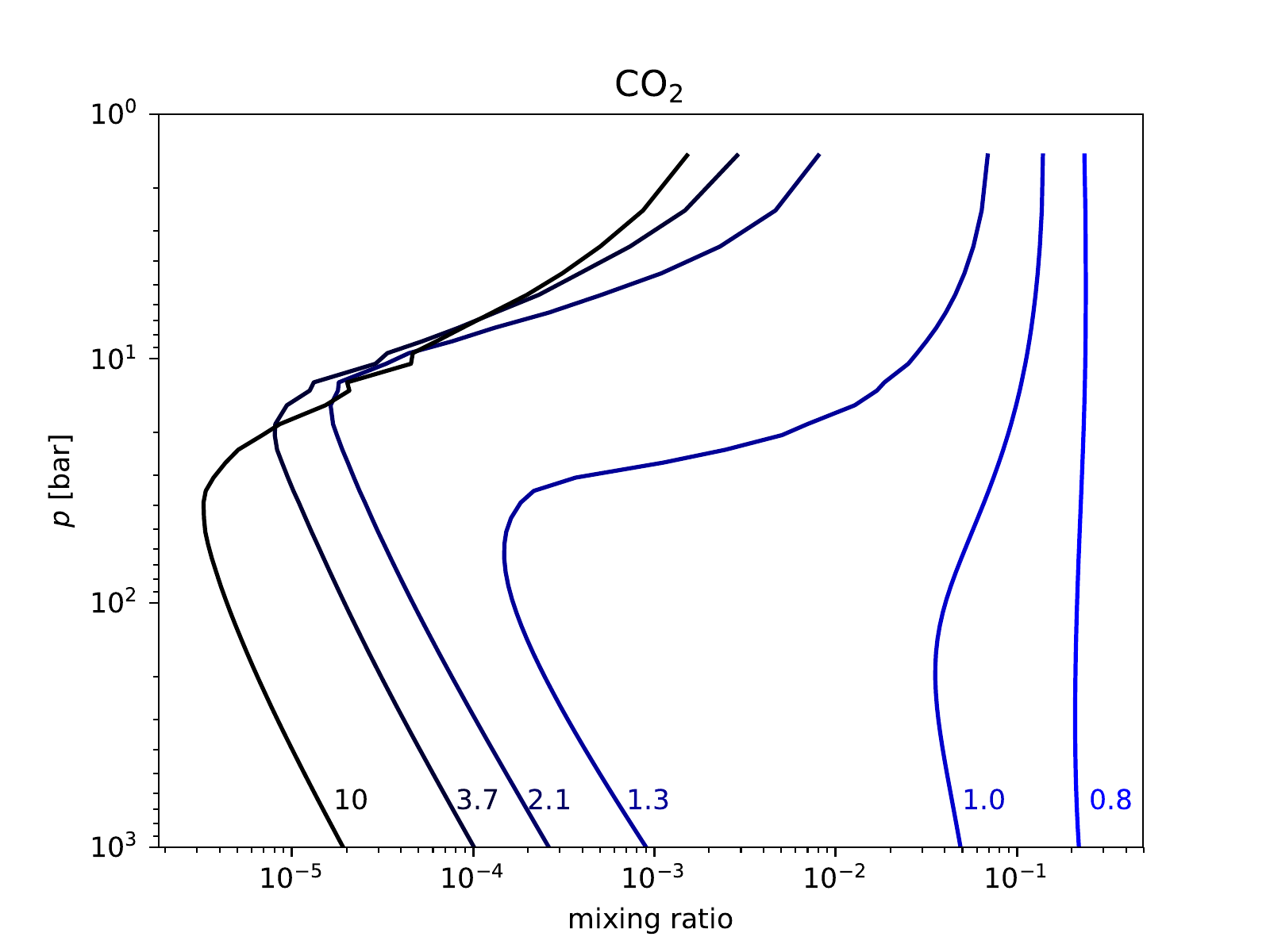}
\includegraphics[width=0.65\textwidth]{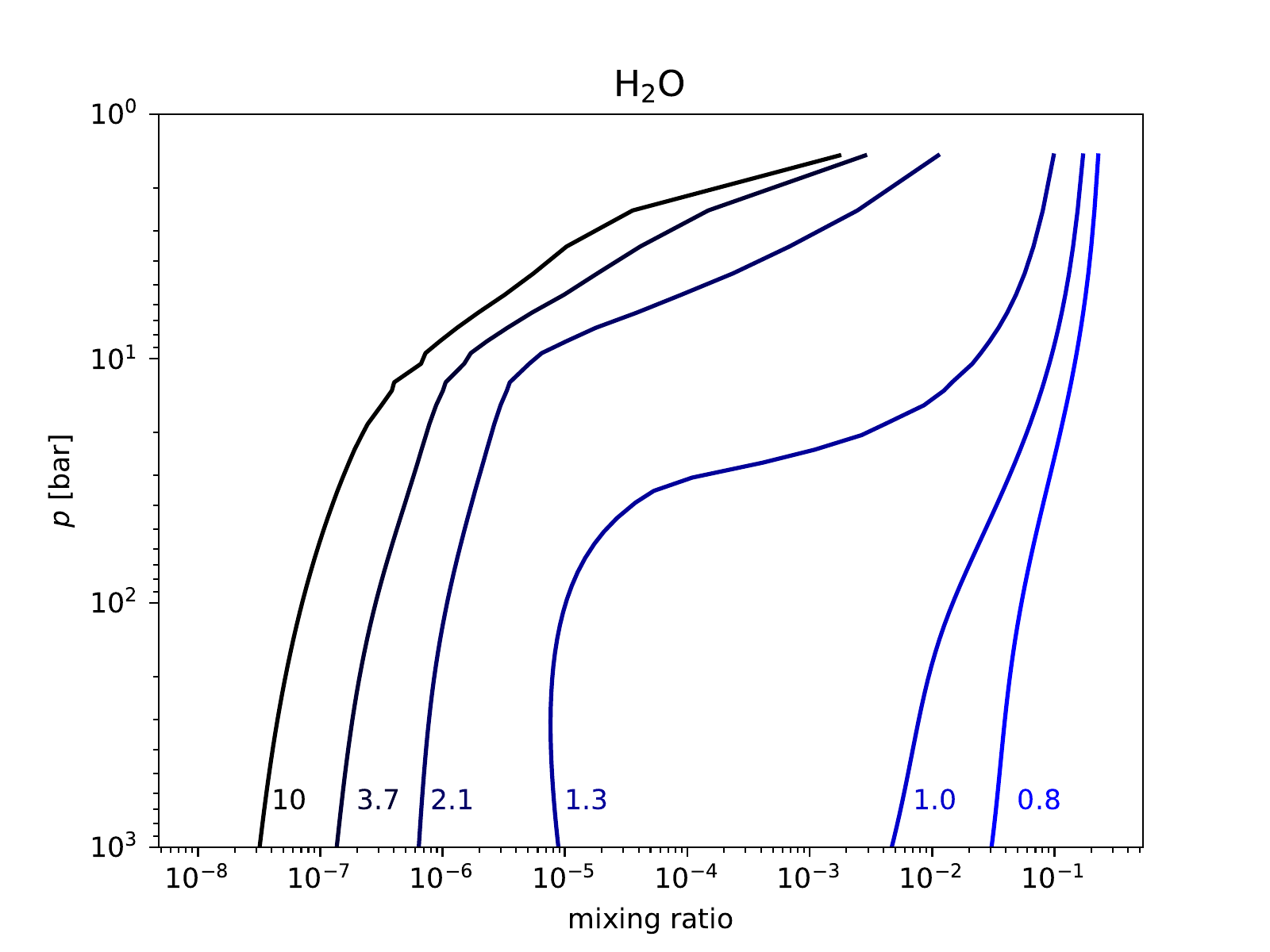}
\caption{The mixing ratios of water, CO and \ce{CO_2} as a function of pressure for various C/O ratios from 0.8 (blue) -- 10 (black), with the values noted next to the relevant curves. \label{fig:water}}
\end{figure}

\begin{figure}
\centering
\includegraphics[width=0.8\textwidth]{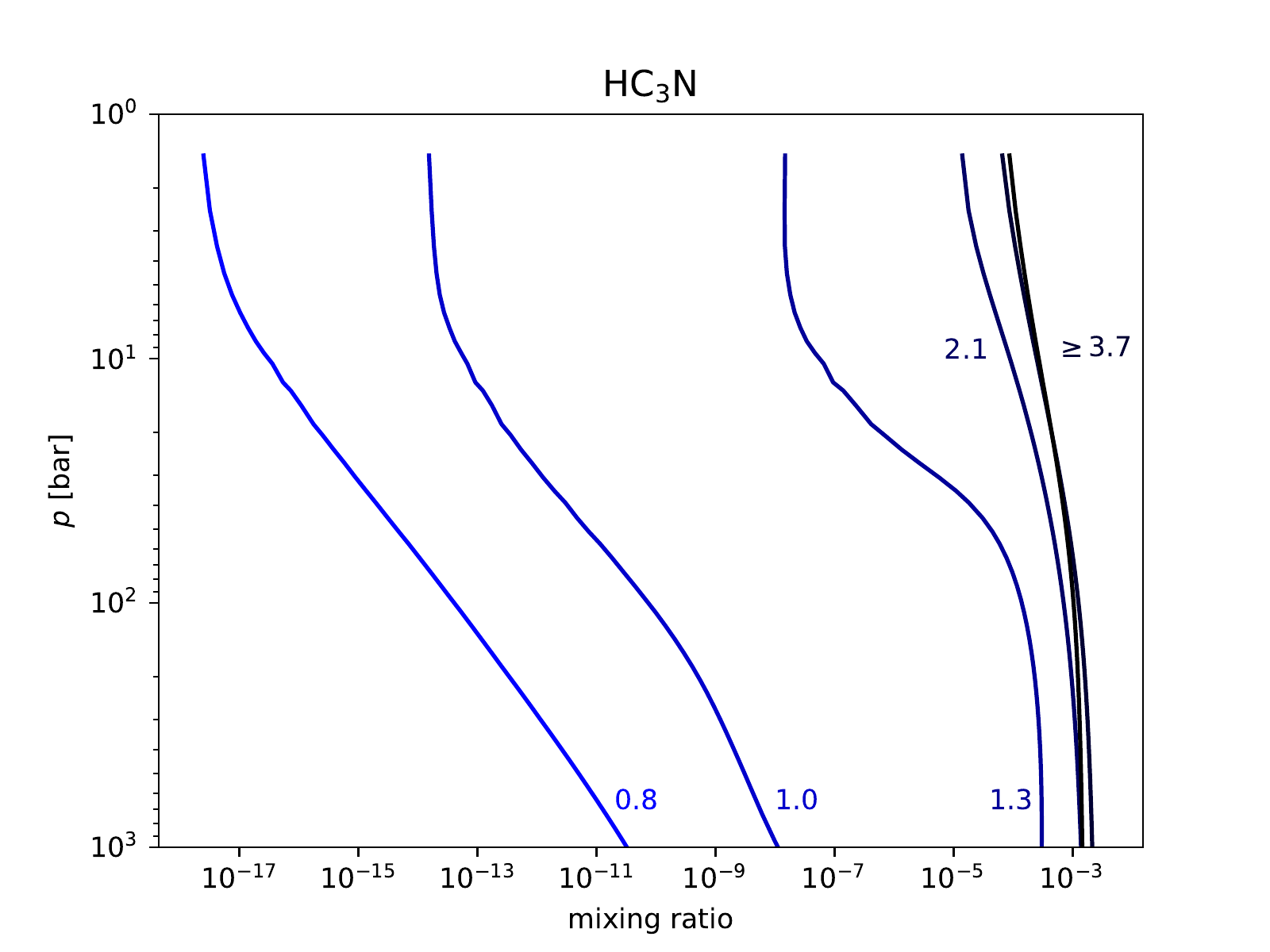}
\includegraphics[width=0.8\textwidth]{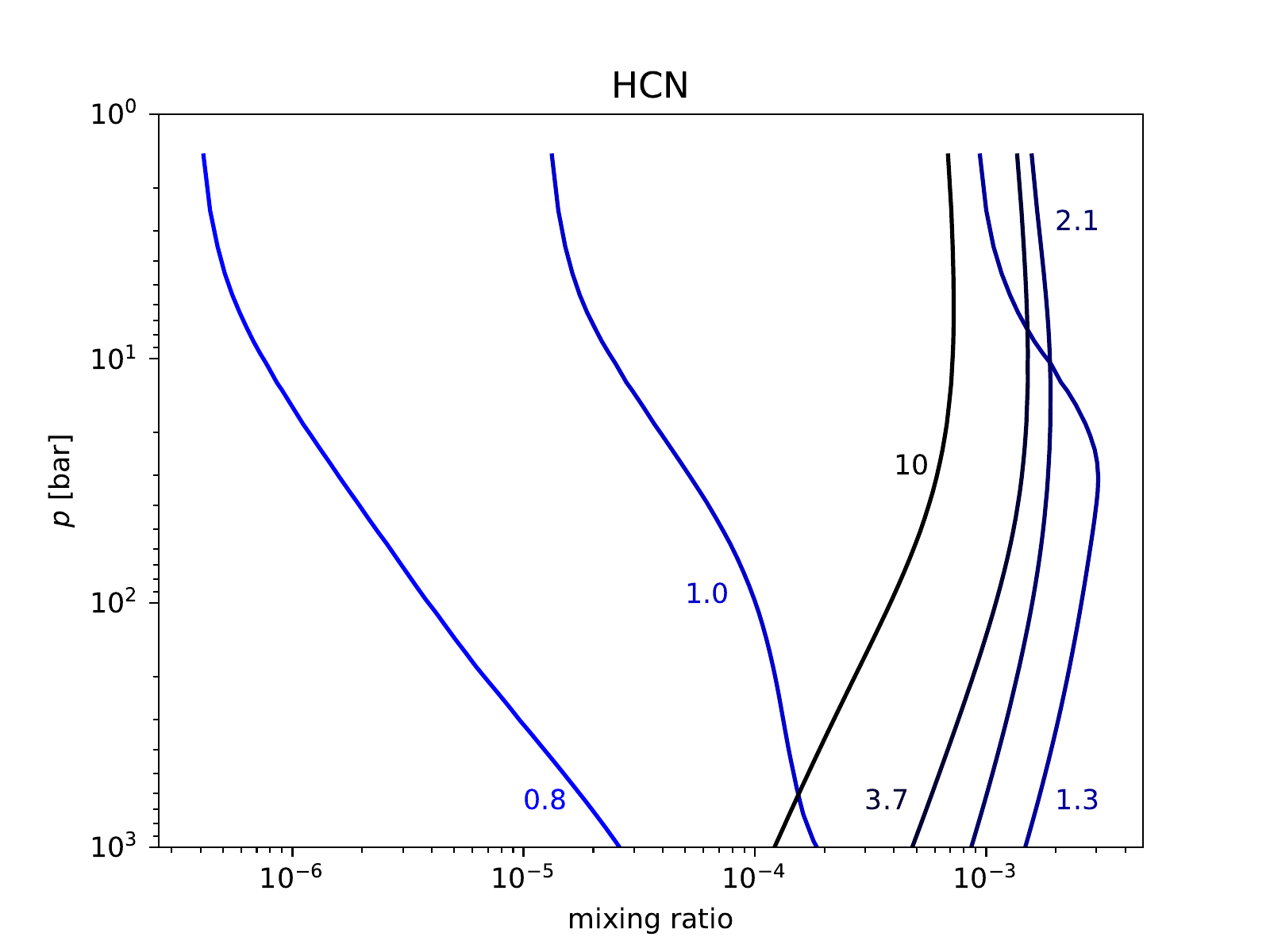}
\caption{The mixing ratios of the nitriles, HCN and \ce{HC_3N}, as a function of pressure for various C/O ratios from 0.8 (blue) -- 10 (black), with the values noted next to the relevant curves. \label{fig:nitrile}}
\end{figure}

The oxygen fugacity ($f_{\rm O_2}$) is signficantly lowered in the presence of excess carbon, because the carbon will rapidly react with the oxygen, binding it up in the forms of \ce{CO_2} and \ce{CO}. Therefore the added carbon has a profound affect on the magma, and results in an ultra-reducing magma with a very low $f_{\rm O_2}$. Added nitrogen does not affect the oxygen significantly. At the same time, decreasing the amount of available carbon results in significantly less HCN and \ce{HC_3N}. The Carbon-poor Nitrogen-poor magma is set by the NiNiO buffer, with $\Delta \ce{NNO} = -1$. To estimate the change in $f_{\rm O_2}$ in the carbon-rich magma from the carbon-poor, we compare the mixing ratios of oxygen as so:
\begin{equation}
\delta f_{\rm O_2} = \dfrac{[\ce{O_2}]_{\rm CR}}{[\ce{O_2}]_{\rm CP}}.
\end{equation}
where $[X]$ is the volume mixing ratio of species $X$, and CR refers to the carbon-rich case, and CP to the carbon-poor case. The $f_{\rm O_2}$ is shown for different C/O ratios in Fig. \ref{fig:fo2}. For comparison, the evidence presented by Yang et al. (2014)$^{Y}$ justifies a global $\log \, \delta f_{\rm O_2}$ of -5 at most. As we discuss in the paper, there may be ways to achieve much lower $f_{\rm O_2}$ in local environments. The effect of this added carbon on other species, \ce{CO}, \ce{CO_2}, \ce{H_2O}, \ce{HCN} and \ce{HC_3N} are also shown in Fig's. \ref{fig:water} and \ref{fig:nitrile}.


\pagebreak


\reftitle{References}
\bibliographystyle{unsrt}





\end{document}